\documentstyle[twoside,fleqn,espcrc2]{article}

\input epsf
\epsfverbosetrue

\title{The status of US Teraflops-scale projects}

\author{Robert D. Mawhinney\address{Department of Physics,
	Columbia University, New York, NY, 10027, USA}
        \thanks{Research sponsored in part by the United States
	Department of Energy} }
       
\begin{document}
\def\thepage{CU--TP--666 \ \ \  hep-lat/9412068}
\thispagestyle{myheadings}
\begin{abstract}
The current status of United States projects pursuing Teraflops-scale
computing resources for lattice field theory is discussed.  Two
projects are in existence at this time: the Multidisciplinary Teraflops
Project, incorporating the physicists of the QCD Teraflops
Collaboration, and a smaller project, centered at Columbia, involving
the design and construction of a 0.8 Teraflops computer primarily for
QCD.
\end{abstract}
\maketitle
\section{INTRODUCTION}
Along with new analytic approaches and algorithmic improvements,
increasing computational power is an obvious avenue for progress in
lattice field theory, particularly QCD.  To achieve a significant
increase in physical volume and a decrease in lattice spacing requires
a large increase in computer power, precipitating a move to machines of
the Teraflops scale.  On such machines, calculations with dynamical
fermions should have statistical errors and control of systematic
errors from finite volume and non-zero lattice spacing comparable to
the best quenched calculations of today.

There are currently three groups in the US doing calculations on
dedicated, multi-Gigaflops scale computers built primarily for QCD;
they are groups at Columbia, Fermilab and IBM.  Currently, the IBM
group has no plans for a next generation machine and the Fermilab group
is investigating options for a machine to replace ACP-MAPS in the near
future.  Over the last few years, the members of the Columbia group
were active participants in the QCD Teraflops Project, a US
collaboration of many tens of physicists working to acquire a
Teraflops-scale machine for QCD.  This project involved a commercial
partner, Thinking Machines Corporation, with a strategy centered on
increasing the floating point power of a commercial machine for QCD.

In spring of 1993, a decision by Thinking Machines Corporation to
discourage our use of the CM-5 platform for this enhanced machine and a
decision by Texas Instruments (TI) not to participate in the floating
point upgrade meant a serious delay in the development of this
project.  At that time, the Columbia group and a few collaborators
began working to design and build a dedicated, inexpensive 0.8
Teraflops machine primarily for QCD.  The Multidisciplinary Teraflops
Project is the name generally given to the continued collaborative
effort with Thinking Machines, with the QCD Teraflops Collaboration as
a major part.  We will focus on the status of these two projects

\section{MULTIDISCIPLINARY TERAFLOPS PROJECT}

The Multidisciplinary Teraflops Project has been seeking to develop
enhanced floating point hardware to be mated with the network that
Thinking Machines Corporation was planning for the successor to the
CM-5.  Another major goal has been the ability to make use of the
existing Thinking Machines software tools.  The QCD physicists working
on this project will provide a major application for the resources
provided by this relatively general purpose machine.  The project also
involves the Laboratory for Computer Science at MIT and Lincoln
Laboratories.

The proposed machine is a 64 bit, MIMD computer with 512 floating point
accelerators.  Each accelerator is attached to a general communications
network, producing a peak speed of 1.6 Teraflops with 250 Gbytes of
memory for a cost of 25 million dollars.  The recent bankruptcy of
Thinking Machines Corporation has added a large measure of uncertainty
to the project.  At this time it is unclear whether Thinking Machines
network technology will reappear in the marketplace, or whether another
commercial network can be used, so we will not be able to detail the
complete machine and instead will concentrate on the architecture of
the the floating point accelerator, which has been under active
development.

The accelerator consists of 16 Power PC 604 RISC processors, each
connected to 32 Mbytes of local memory (either DRAM or synchronous
DRAM) through a memory controller.  In turn, each memory controller is
attached to an industry standard, 64 bit wide,  PCI bus with a
bandwidth of 422 Megabytes/sec at 66 MHz.  The Power PC processor
supports multiple cache lines for efficient prefetching of data from
the local memory, has 32 registers and a 32 kilobyte data cache.  Each
of the Power PC's has a peak speed of 200 Mflops, giving the
accelerator board a peak speed of 3.2 Gflops at an expected cost of
\$28,000.

The architecture has the memory controller for each processor capable
of sending data to any other processor in the machine.  The memory
controller may only be able to transfer sequential blocks of data, ie.\
no block-strided transfer capability, which requires the occasional
involvement of the Power PC to rearrange data before a transfer.

Simulations of this design have given estimates for the floating point
performance of the application of the staggered Dirac operator.  The
simulator has been developed by members of the collaboration and uses
their understanding of the cache structure of the Power PC to model its
performance.  Their results show sustained performance of about 1.15
Teraflops.  The most heavily loaded bus is the one connecting the
floating point accelerator to the 512-node network, and it is loaded at
slightly below 50\% of its capacity.  These simulations do not include
the global sums necessary in the staggered conjugate gradient.

There are currently efforts underway to check the performance of the
accelerator on other applications, an important question given the
multidisciplinary use expected.
\pagenumbering{arabic}
\addtocounter{page}{1}
\section{THE COLUMBIA 0.8 TERAFLOPS COMPUTER}

In the spring of 1993 a group led by physicists at Columbia began work
on a computer capable of Teraflops scale computing for a few million
dollars.  We have pursued a strategy which makes as much use as
possible of commercially available components arranged in an efficient
form for QCD calculations.  To this end, we have been working to design
and construct a 0.8 Teraflops, 32 Gbyte, largely 32 bit precision, MIMD
computer based on a four dimensional, $16^3 \times 4$ array of 16,384
Digital Signal Processors (DSP's) with 25 or 50 MHz bit-serial
nearest-neighbor communication between processors\cite{nhc}.  The total
price is 3 million dollars.

The members of the collaboration are
\begin{itemize}
\item { \bf Columbia University:} \\
   Igor Arsenin, Dong Chen, Norman H. Christ, Chulwoo Jung,
   Adrian Kahler, Roy Luo, Robert D. Mawhinney, and Pavlos Vranas
\item { \bf Columbia University Nevis Laboratory:} \\ 
   Alan Gara and John Parsons
\item { \bf Fermilab:} \\ 
   Sten Hanson
\item { \bf SCRI at Florida State University:} \\ 
  Robert Edwards and Tony Kennedy
\item { \bf The Ohio State University:} \\ 
  Greg Kilcup
\item { \bf Trinity College, Dublin:} \\ 
  Jim Sexton
\end{itemize}

\begin{figure*}[t]
\epsfxsize=5.0in
\epsfbox[0 300 500 550]{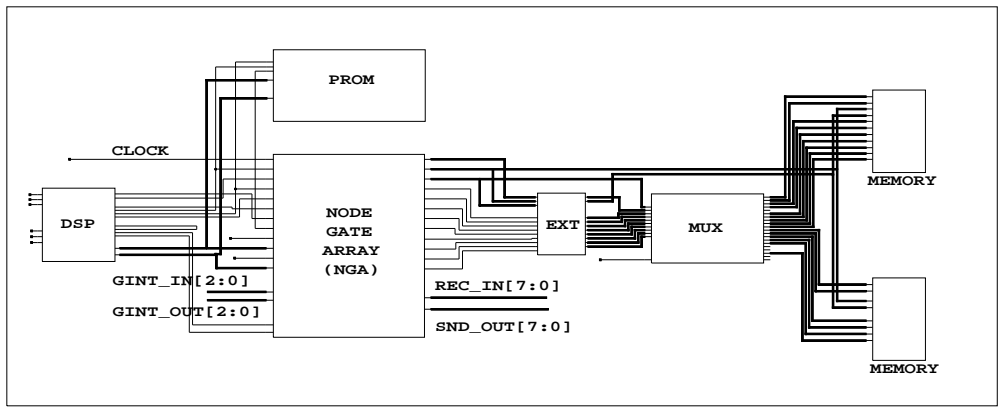}
\caption{The schematic of our processing node as it appears in the
ViewLogic system we have used in design and testing of the node gate
array (NGA).  The blocks labeled DSP, PROM and MEMORY include models
from Logic Modeling, Inc., which accurately reflect the detailed
behavior and timings of the commercial components we will use.  The DSP
model runs DSP programs written in C or assembly.}
\label{fig:node}
\end{figure*}

The basic processing node of our machine (Figure \ref{fig:node})
consists of a DSP, 1/2 Mword of memory (five chips) and a custom-made
chip we call the node gate array (NGA).  This seven chip processing
node will fit on a double-sided printed circuit board of size about 1.8
inches by 2.7 inches, has an expected cost of \$140 and will use 2--3
watts of power.

16,384 of these nodes gives a 0.8 Teraflops computer that uses around
50 kilowatts of power (the current Columbia 256-node computer uses 30
kilowatts) and contains roughly 110,000 chips (the 256-node computer
has 80,000).  For the DSP and DRAM chips we are using, the hardware
failure rate is a few tens in a billion hours of use.  Therefore, even
with $\sim 10^5$ chips and $\sim 10^4$ hours in a year, we expect to
have a few tens of failures per year, or roughly one every few weeks
(after five years of running, the 256-node machine has one every few
months).  This low failure rate removes the need for fault-tolerant
software and allows us to operate the machine only when there are no
hardware failures.
\subsection{Processor}
To achieve Teraflops scale computing for a price of a few million
dollars, processing power alone has to cost around one dollar for a
Mflops.  Our answer to this constraint has been to focus on Digital
Signal Processors.  Today, DSP's are a commodity that appear in modems,
CD players, etc.  They are essentially microprocessors where the
emphasis has been placed on arithmetic power (floating point power in
our case) as opposed to, for example, context switching efficiency,
which is necessary for a microprocessor handling a multitasking
environment like a UNIX workstation.

We have chosen to use a Texas Instruments TMS320C31 DSP, a single
precision, 50 Mflops DSP which is expected to be available in quantity
for around \$50.  This chip uses about a watt of power and is available
as a surface mount component.  The CPU can perform two operations in a
single 25MHz cycle; the instruction set includes parallel load-store
operations as well as parallel multiply-add's.  In addition to
16 registers, there is a 2 kiloword memory inside the DSP which can
supply two of the source operands for the parallel operations.  The DSP
also has an internal DMA, which can be used, for example, to load data
into the on chip memory while the CPU is doing calculations.  The DSP
can also boot over a serial line, which precludes the need for 16,384
PROM's.
\subsection{Architecture}
We have faced two main architectural challenges in turning a collection
of processors into an efficient parallel machine.  They are:  to
develop a simple, efficient, inexpensive communications network for QCD
and to balance the speed of the DSP with the memory bandwidth available
from DRAM.  In this section, we will describe our solutions to these
challenges.
\subsubsection{Communication network}
Given the large number of processors in our machine, a completely
general communication network, such as a crossbar switch, is either
prohibitively expensive or too slow.  However, most of the
communication for known QCD algorithms involves only nearest-neighbor
processors (with the exception of the global sums described below).
Consequently, we have chosen a bit-serial, nearest-neighbor
communication network to connect the $16^3 \times 4$ array of
processors.

Analyzing the communication bandwidth requirements of the staggered
conjugate gradient, for the demanding case of $2^4$ lattice sites per
node, gives the desired frequency of the serial links.  Assuming
complete overlap of communication and computation, simultaneous
bit-serial communication between nearest neighbors in all directions (8
for four dimensions) gives sufficient bandwidth with 25MHz serial
links.  This overlap can be quite efficiently achieved for the
staggered conjugate gradient, but in order to have extra bandwidth when
the overlap is not complete, we have designed our serial communications
network to run at either 25 or 50 MHz.

Each node contains a serial communication unit (SCU) controlling the
eight autonomous nearest-neighbor links connected to the node.  The
communication between two nodes at the ends of a serial link is
completely independent of any other serial connection and serves to
synchronize the entire machine.  The link level communication is word
(32 bit) oriented and contains a simple protocol which signals the
beginning of a transmission, the word itself, a parity bit and
concludes with an acknowledgement packet from the receiving neighbor.
Single bit errors in the transmission result in an automatic retry at
the hardware level;  these errors are transparent to the user, except
for a counter logging their total number.  Also, each of the eight
communications controllers on a node has its own DMA, which includes
the ability to do block-strided addressing of either the data sent or
the data to be stored into local DRAM.

A major problem in large parallel machines is efficient performance of
global sums.  With our bit-serial network, the 40 cycle (including
parity and handshaking) latency for the arrival of a given word at its
neighboring node makes implementing a global sum by passing the sum
from neighbor to neighbor through the machine very inefficient.  To
avoid this latency, we have included in the SCU various {\em
pass-through} modes where the SCU itself takes action on the incoming
bit stream as each bit arrives.  The result is immediately sent out
with a latency of one cycle per node.

\begin{figure}[htb]
\epsfxsize=2.5in
\epsfbox[0 310 450 532]{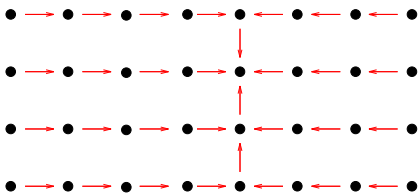}
\caption{An example of the serial communication paths for a
global sum or max operation.}
\label{fig:pass_through}
\end{figure}

We have implemented {\em max}, {\em sum} and {\em broadcast} operations
as our possible pass-through modes.
\begin{enumerate}
\item  In a {\em max} or {\em sum} operation, input is taken
  from up to seven of the eight serial links to the node, plus
  local memory.  The data from these inputs has a max or sum performed
  on it, with the result sent out via a single serial link.
\item In a {\em broadcast} operation, input from a single serial
  link is stored in local memory and sent out over a chosen set of
  serial links.
\end{enumerate}
An example of the data paths for a global sum or max operation is shown
in Figure \ref{fig:pass_through}.  The processor with all arrows
pointing toward it finally holds the required result.  Six cycles after
the sum starts, the first bit from the processor in the lower left
corner arrives.

An additional feature of the pass-through modes is the ability to do
arbitrary precision global sums efficiently.  The SCU can continue to
add the results of a previous 32-bit pass-through sum to the next
pass-through sum.  Since precision is most important in the global
sums, we have the ability to easily insure that our global sums are
accurate.

The SCU control registers are memory mapped into the DSP address
space.  A send operation to one of the eight neighbors requires the DSP
to write to two SCU control registers. One write programs the DMA for
this transfer with information about the number of blocks of data, the
size of each data block and the offset between blocks.  The second
gives the starting address of the data to be sent and starts the send.
Programming the receiving node is similar and also takes two writes.
\subsubsection{Circular buffer}
To achieve a balance between the DSP speed and the DRAM bandwidth, we
have implemented a fetch-ahead cache, which we call the circular buffer
(CIRBUF).  The CIRBUF contains 32 words of memory, which it manages
according to how it is programmed, and it can remember five different
programming modes concurrently.  Programming modes are generally set
before entering a particular subroutine and fall into two general
categories:
\begin{itemize}
  \item {\bf one wait-state modes:} \ \ In these modes, every DSP
    read from memory takes at least two 25 MHz cycles.  The CIRBUF is
    programmed to try and stay a few words ahead of the address of the
    last DSP read, to achieve as many two cycle reads as possible.  The
    programmer does not need to worry about the pattern of fetches from
    memory; the CIRBUF will respond as quickly as possible no matter
    how the program jumps around.  This mode is useful
    when running a general C program, for example.
  \item {\bf zero wait-state modes:} \ \ In these modes, the CIRBUF can
    provide a word from memory every 25 MHz cycle, after an initial
    delay of three cycles for the first word.  To achieve this
    performance, the CIRBUF must {\em know} it already has any word it
    will be asked for.  It can do this, since the programming mode
    contains information about the maximum difference in the addresses
    between consecutive requests.

    For example, to read sequentially through an SU(3) matrix, as is
    necessary for a matrix times vector product, you will never jump
    ahead by more than one word beyond the previous word fetched.  For
    the multiplication of the hermitian conjugate of this matrix times
    a vector, you would commonly skip ahead by five words and hence
    would use a different mode for the circular buffer.
\end{itemize}

Physical memory appears multiple times in the address space of the
DSP.  Reading the same physical memory under the control of different
programming modes of the circular buffer is accomplished by reading a
different image of memory.  This involves the addition of a constant
offset to the addresses of the data to be fetched.

The CIRBUF gives an obvious large speed increase compared to the three
cycle reads that could be sustained from DRAM without its presence.  We
have also found it quite straightforward to write programs, where
performance is important, using the zero wait-state modes

\subsection{Node gate array}

\begin{figure*}[t]
\epsfxsize=5.0in
\epsfbox[0 250 500 550]{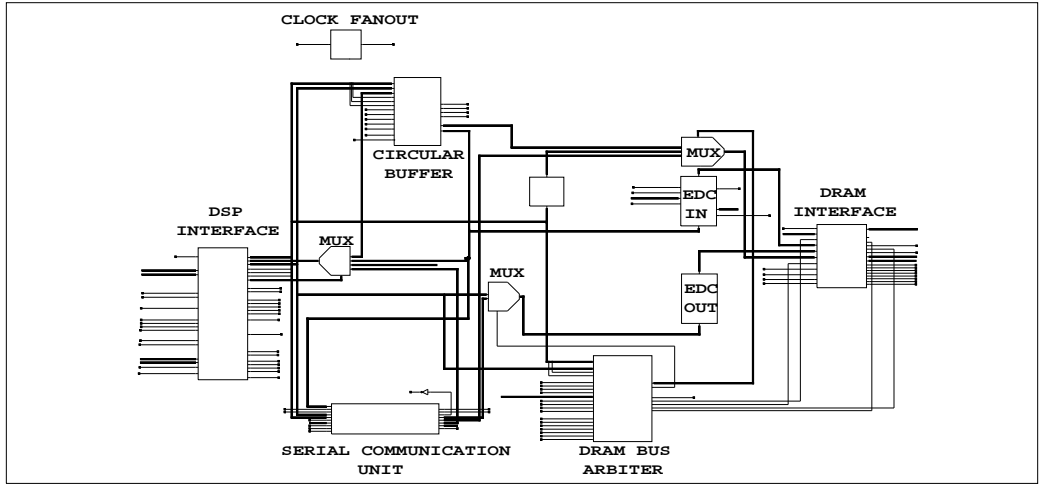}
\caption{The top-level schematic of the NGA.  Each block (DSP INTERFACE,
etc) contains hierarchical layers of ever-more-detailed schematics.}
\label{fig:nga}
\end{figure*}

The SCU and CIRBUF discussed in the previous sections are implemented
in an application specific integrated circuit (ASIC), called the node
gate array (NGA) (Figure \ref{fig:nga}), which we have designed.  The
NGA also contains an interface to DRAM, which handles DRAM refreshes
and incorporates a double-bit error detection and single bit error
correction unit; a module to arbitrate DRAM access between the SCU, the
CIRBUF and direct DSP reads; registers to latch and send global
interrupts; and circuitry for clock synchronization and other boot-time
issues.

This is the one customized chip required for this project.  It will
have 208 pins and be about 1.25 inches on each side.  The architecture
has been implemented using a high-level hardware description language
described elsewhere in these proceedings \cite{igor}, which is then
turned into schematics by a compiler-like tool called a synthesizer.
We have a complete schematic representation for our NGA and are ready
to send it to the engineering staff of an ASIC manufacturer for final
testing and production.
\subsection{Complete machine}
We will attach 64 nodes of our machine to a motherboard using SIMM-like
edge connectors, making it easy to replace a failed node\cite{nhc}.  On
each motherboard, there will also be two SCSI ports for I/O to the host
SUN workstation and a parallel disk/tape array.  With over 5 Mbytes/sec
bandwidth per SCSI port and 256 motherboards, the bandwidth to 256
disks is easily 1 Gbyte/sec (We expect to begin with 32 or 64 disks.).

Each motherboard will have a $ 2 \times 2 \times 4 \times 4 $ array of
processors.  Eight motherboards will be placed together in a commercial
crate and 4 crates, separated by heat exchangers, will be stacked into
a rack.  Cabling can be done so that no cable is longer than about 6
feet.
\subsection{DSP Tools}
The DSP comes with a complete software development environment from TI,
including assembler, C compiler, and a graphical debugger.  Application
code is also available from a TI bulletin board.  As an example, when
we needed to rearrange the compiler output for use as input to our NGA
simulations, a subroutine was added to a bulletin board program.

In addition, we have a number of commercial DSP boards plugged into a
SUN and various PC's, permitting us to program on real hardware.  In
particular, one of the DSP boards (from DSP Research) has a SCSI port
driven by the DSP, allowing development of the SCSI software for the
connection between the 0.8 Teraflops machine and a SUN workstation.

Finally, we also have an emulator pod from TI which allows an external
DSP to be completely controlled from a personal computer.  This will be
extremely useful when we have our first prototype hardware.  The user
controls the external DSP through a graphical interface that is the
same as the interface to the simulator and debugger.
\subsection{Software}
We have gained considerable experience programming the DSP, using the
various DSP boards.  The TI software development tools are very robust
and complete.  We have also done extensive programming of the DSP and
NGA assembly together, using the ViewLogic environment and the Logic
Modeling model of the DSP\cite{igor}.  This is part of the development
and testing process for the NGA and is a gate level simulation of the
NGA.

We currently have both a staggered and Wilson fermion conjugate
gradient code running on our node, using the CIRBUF in zero wait-state
modes and the SCU, without global sums.  These programs have not only
allowed us to gain experience in programming the machine for high
performance, but have also caught many errors in our NGA design.  The
programs involve assembly-code subroutines to do the application of the
Dirac operator, with C programs performing initializations and high
level control.  (Including comments, the staggered Dirac operator
assembly code is about 1,000 lines long.)

The staggered conjugate gradient for a $2^4$ lattice, running on actual
DSP hardware with SRAM (so there are no memory bandwidth delays),
sustains 56\% of peak speed.  This same program on our node, using DRAM
and the CIRBUF sustains 46\%, showing that the CIRBUF is making the
DRAM appear fast.  A quick modification of the program to use the
serial communication network (so we could test the NGA) achieves 30\%,
which should go to at least 40\% with a little tuning.

The Wilson conjugate gradient code uses the DMA to load the DSP's
internal memory and sustains over 30\% on a $2^4$ lattice.  This
performance will increase as the lattice volume per node increases.  In
addition, our simulator currently is running with an artificially rapid
refresh rate for the DRAM, so we can look for errors in the DRAM
controller.  This degrades all our performance figures.

These results give a lower bound on the performance we can expect on
the real machine.  Our programming experience has also given us ideas
about more efficient programming schemes.
\subsection{Physics goals}
The primary physics goals are quite clear; to do QCD simulations with
as much control over errors as possible.  Comparison of dynamical
fermion and quenched calculations should reveal the accuracy or
breakdown of the quenched approximation.

A second major goal of this project is to archive the configurations we
produce for use by others.  To this end we are implementing an
archiving system on our current 256-node machine, which will grow to
hold the configurations of the 0.8 Teraflops machine.  These
configurations will provide the community with a common set of lattices
with known properties (lattice scale, hadron masses, etc.) and allow
extraction of the full spectrum of physics available.
\section{SUMMARY}
The Multidisciplinary Teraflops Project has specified an architecture
for the accelerator node, but must find a network to host it.  About
\$1 million has been spent, with about \$3.2 million needed for the
completion of a few boards.

The Columbia 0.8 Teraflops computer has its NGA completely designed and
the first hardware is expected to be ready in a few months.  Funding
has been acquired for a 512-node prototype with a request for funding
of the full machine to be made upon successful operation of the
prototype.
\\
\\
{\bf Acknowledgements:} I would like to thank Estia Eichten, Andreas
Kronfeld and Don Weingarten for discussions and Richard Brower and John
Negele for providing detailed information about the Multidisciplinary
Teraflops Project.

\end{document}